\documentclass[
    ,final            % use final for the camera ready runs
  ]
  {aipproc}

\layoutstyle{6x9}

\usepackage{amssymb} % Need some of symbols including <~

\newcommand{\etal}{et al}   			% et al in appropriate format, the following . 
\newcommand{\lapprox}{\ensuremath{\lesssim}}

\begin{document}

\title{Status of Direct Searches for WIMP Dark Matter}

%                \texttt{http://www.aip..org/pacs/index.html}>}

\author{Richard W. Schnee}{
  address={Department of Physics, Case Western
Reserve University, Cleveland, OH  44106, USA}
}

\begin{abstract}

Astrophysical observations indicate that about 23\% of the energy density of the universe
is in the form of non-baryonic particles beyond the standard model of
particle physics.  
One exciting and well motivated candidate is the lightest supersymmetric partner particle (LSP),
which could be a weakly
interacting massive particle (WIMP) left over from the Big Bang.
To determine that the LSP is the dark matter, it is necessary both to measure the particle's properties at an accelerator and to detect the particle in the galaxy directly (or indirectly).
Direct detection of
these particles requires sophisticated detectors to defeat much 
higher-rate backgrounds due to radioactivity and other sources.
Promising techniques identify individual interactions
in shielded fiducial volumes and distinguish nuclear-recoil 
signal events from electron-recoil backgrounds, based on the timing, 
energy density, and/or the division of the energy into
signals of ionization, scintillation, or phonons. 
I  review the techniques of the dozens of experiments
searching for WIMPs and summarize the most interesting results and prospects
for detection.

\end{abstract}

\maketitle

%%%%%%%%%%%%%%%%%%%%%%%%%%%%%%%%%%%%%%%%%%%%
%% MAINMATTER
%%%%%%%%%%%%%%%%%%%%%%%%%%%%%%%%%%%%%%%%%%%%

\section{WIMPs and Supersymmetry}

Observations of 
the cosmic microwave background, galaxy clusters, and distant supernovae are all consistent with the conclusion that the universe is geometrically flat, 
with most of the energy density of the universe in the form of some dark energy or cosmological constant, and about 30\% in the form of dark matter~\cite{wmap2006params}. 
Measurements, including ones of the primordial abundance of light elements~\cite{TytlerBBN2001}, indicate that 80\% of this dark matter is some form of cold, non-baryonic dark matter.  A strong candidate is weakly interacting massive particles (WIMPs)~\cite{WIMP}, of which the strongest candidate is the lightest supersymmetric partner particle, possibly the neutralino~\cite{Reviews}.  

WIMPs can potentially be detected by by three complementary methods.  
They may be produced and detected (indirectly) at accelerators such as the Large Hadronic Collider.
Relic WIMPs may be detected indirectly via their annihilation products using gamma-ray telescopes, neutrino telescopes, or satellites~\cite{indirect}.
Finally, relic WIMPs may also be detected directly when they scatter off nuclei in terrestrial detectors.  
Because WIMPs move at slow, galactic velocities, they tend to cause small momenta transfers 
and hence interact coherently with the entire nucleus~\cite{lewinsmith}.  
Spin-independent interactions therefore scale as the square of the atomic mass of the target and hence dominate for most models over spin-dependent interactions (which are neglected in this review). 

\begin{figure}
  \includegraphics[height=.4\textheight]{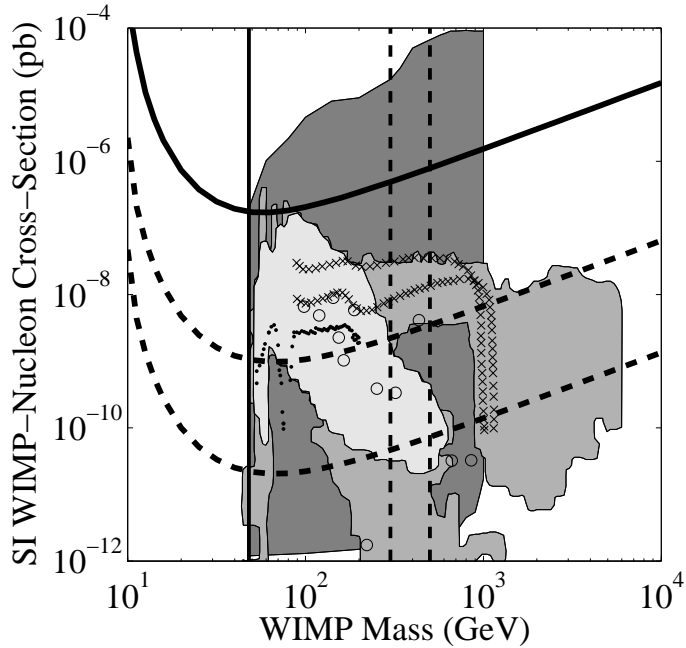}
  \caption{Supersymmetry reach of direct-detection experiments and colliders.
  Allowed parameter space in the plane of spin-independent WIMP-nucleon cross section versus
  WIMP mass is shown for general MSSM models~\cite{Kim02} (dark grey filled region; the cut-off at high WIMP masses is artificial),
  for more constrained, mSugra models~\cite{Baltz04} (medium grey filled region),
 for mSugra models with constraints from the muon g-2 measurement applied (light grey region),
 and for yet more constrained split-Supersymmetry models ($\times$~\cite{Giudice04} and dots~\cite{Pierce04}).  The post-LEP supersymmetry benchmark points~\cite{Battaglia04} are shown as circles.
 The most sensitive current direct detection experiment, CDMS~II~\cite{r119prl} (solid curve) is starting to probe these regions,
based on 20\,ab$^{-1}$ %twenty inverse attobarns 
effective luminosity.
 Future experiments with negligible backgrounds and $\sim$25\,kg ($\sim$1000\,kg) Ge, Xe, I, or W 
would %extend 
reach to the upper (lower) dashed curve.
 Roughly $4\times$ higher mass in Ar or $8\times$ in Ne could provide the same reach.
Current collider limits (solid vertical line) exclude WIMP masses $\lapprox45$\,GeV,
and the Large Hadronic Collider (left dashed vertical line) and International Linear 
Collider (right dashed vertical line) will provide reach to masses up to 500\,GeV.
  }
  \label{susy}
\end{figure}

Figure~\ref{susy} shows 
the complementarity of direct detection with colliders at probing
the predicted supersymmetry parameter space under different theoretical frameworks.
Direct detection has only started to probe this parameter space, which includes
 many model frameworks concentrated at cross sections from $10^{-8}$\,pb to $10^{-10}$\,pb.
Future direct-detection experiments, increasing in mass and background rejection, should extend their sensitivities down to cover this range. 
It is worth noting that direct detection experiments are sensitive 
%to WIMPs of all masses, 
even to WIMPs of very high masses, but are limited by the WIMP-nucleon interaction strength. 
In contrast, accelerators are sensitive essentially to all WIMP-nucleon interaction strengths, but are limited in their mass reach.  
Both accelerators and direct detection are therefore sensitive to some models to which the other is not.  More excitingly, the two methods have significant overlap where both should be able to detect WIMPs, greatly increasing what we can learn about them, and this overlap occurs for the part of parameter space most densely populated by theories. For example, although we can learn much about WIMPs at colliders, it is impossible to determine at colliders if a WIMP is stable.
We can demonstrate that the WIMP is dark matter only if we can identify the same particle in our galaxy by direct or indirect detection. 
Furthermore, the combination of information from  direct detection and colliders may constrain 
the particle properties significantly better than colliders alone~\cite{baltz_lcc}.

\section{Direct Detection of WIMPs}

The expected rate and spectrum of recoil energies depends on how often and with what velocities WIMPs pass through the terrestrial detectors, which in turn depends on how WIMPs are distributed in the galaxy.
%'s dark matter halo.  
Under standard assumptions the WIMP distribution is Maxwellian~\cite{lewinsmith}, 
resulting in an expected energy spectrum that is exponential with typical energies $\lapprox 50$\,keV.  The expected interaction rate, dependent on the WIMP-nucleon cross section and the local WIMP number density, is $<1$ event per kilogram of target material per week. With such a small event rate, it is a daunting task to search for a WIMP interaction amongst the background interactions from natural radioactivity. It is critical both to maximize the total detector mass to make the WIMP signal as large as possible, and to minimize the rate of background interactions from other particles. 

There are about 25 active WIMP-search experiments worldwide 
(for a more detailed review, see~\cite{gaitskellReview2004}). 
These experiments reduce backgrounds both 
through extensive shielding underground and by rejecting events 
%that are 
more likely 
due to the radioactive backgrounds that result in photons, electrons, or alpha-particles.  For example, WIMPs interact so weakly that they never interact more than once in a detector, 
allowing
experiments to reject multiple-scatter events.
Similarly, WIMPs interact uniformly throughout a detector, so it pays to cut interactions near detector surfaces, which tend to have higher backgrounds.  Most significantly, WIMPs tend to interact with an atom's nucleus, while the dominant radioactive backgrounds interact with the electrons, so experiments that 
discriminate between interactions causing an electron to recoil and those that cause a nuclear recoil can reject virtually the entire radioactive background. 

There are three ways to discriminate between electron recoils and nuclear recoils.  First, the energy density produced by nuclear recoils is much higher than that of electron recoils, so some experiments are immune to electron recoils because the energy they deposit is not dense enough to trigger. 
In addition, the division of energy and pulse timing may be different for electron recoils than for nuclear recoils. 
Depending on the material, recoil energy may be converted into light, ionization, and/or phonons.  Some WIMP-search experiments detect only light, but discriminate nuclear recoils from electron recoils due to the timing of the signal. Other experiments detect two of the forms of energy.
The relative amount of energy in each of the two forms is different for nuclear recoils than for electron recoils, so 
calculating the ratio of the energy in one form to the energy in the other
allows background discrimination.

\subsection{Cryogenic Detectors}

\begin{figure}
%  \includegraphics[height=.27\textheight]{0503265EDELWEISS.eps}
 %  \includegraphics[height=.33\textheight,origin=c,angle=270]{CRESST.eps}
%\begin{center}
%\begin{minipage}[c]{1.25in}
%\centering
% \includegraphics[height=.27\textheight]{0503265EDELWEISS.eps}
%\end{minipage}
 %\begin{minipage}[c]{1.25in}
%\centering
%\includegraphics[height=.33\textheight,angle=270]{CRESST.eps}
%\end{minipage}
%\end{center}
 \includegraphics[height=.28\textheight]{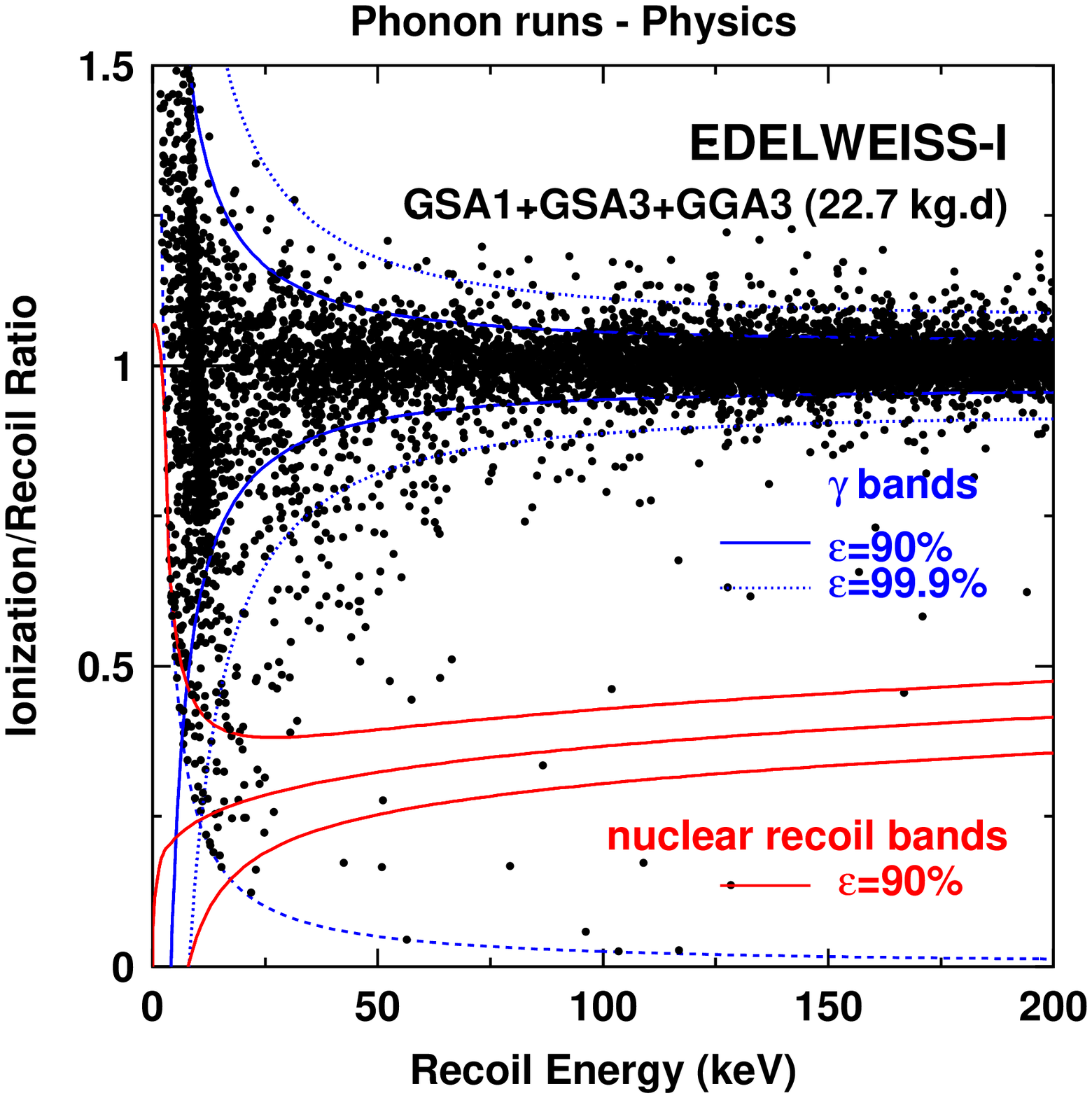}
\rotatebox[origin=br]{-90}{\includegraphics[height=.39\textheight]{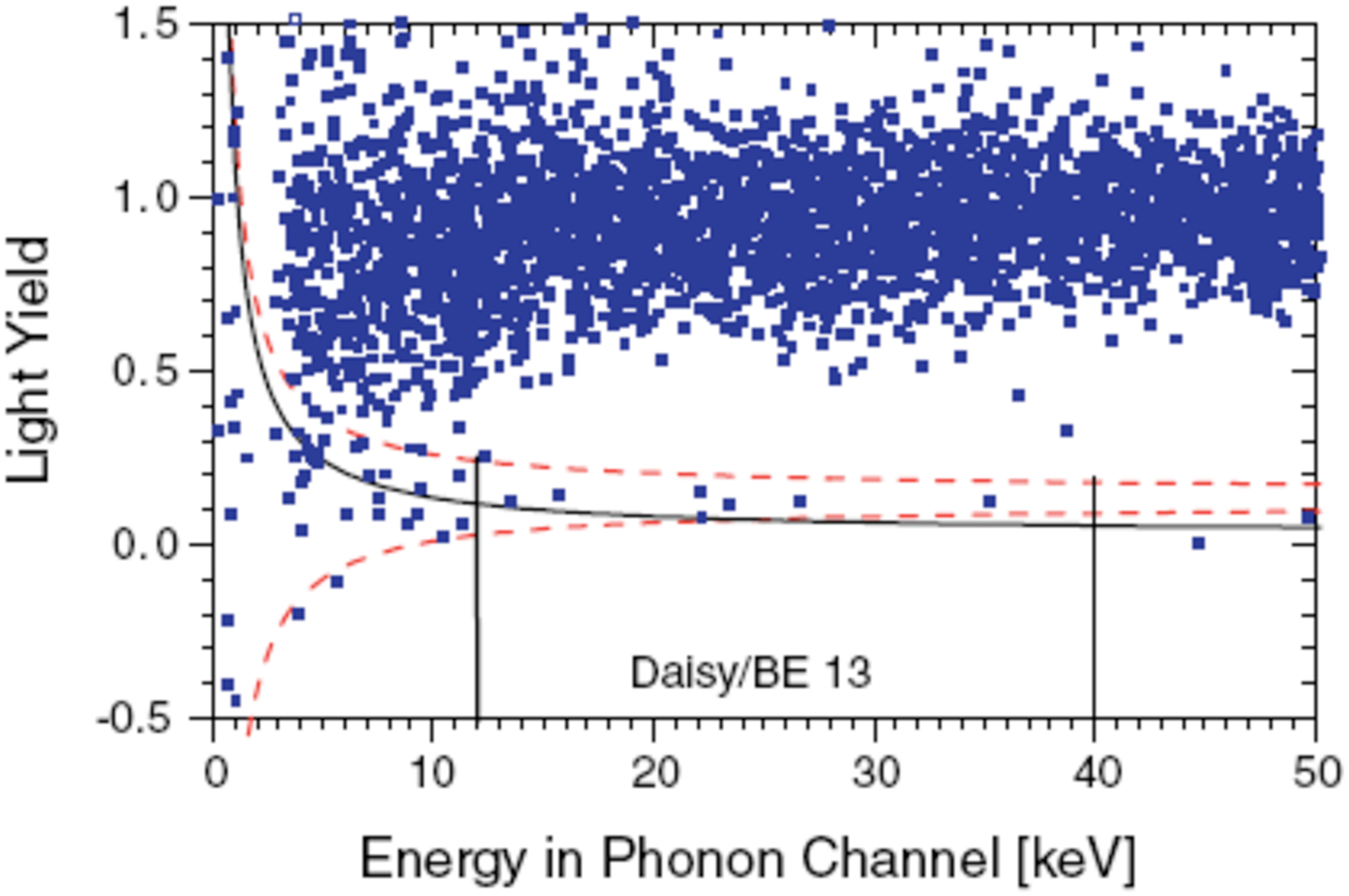}}
 \caption{{\it Left}: Ratio of ionization signal to recoil energy as a function of recoil energy for 
 about one third the exposure of the EDELWEISS-I experiment.  The three events 
 $>50$\,keV within the nuclear-recoil bands (lower set of 3 solid curves) are likely due to neutrons, while the many events in the band at low energy are likely caused by surface electron recoils.
 Figure from~\cite{edelweiss05}.
 {\it Right}: Ratio of light signal to recoil energy as a function of recoil energy for a 10\,kg-d exposure of CRESST's better detector~\cite{cresstfigure}.   
 Events within the oxygen nuclear-recoil band (dashed region) are consistent with neutron interactions.
The lack of events between 12--40\,keV consistent with WIMP interactions on tungsten (below the solid curve, essentially consistent with no light)
 is used to set upper limits on the WIMP-nucleon cross section.
 }
 \label{cryo}
\end{figure}

The EDELWEISS experiment uses thermistors attached to Ge crystals  at cryogenic temperatures 
(20\,mK) to measure phonons, in addition to measuring ionization using a small applied electric field.  
Calibrations with a
neutron source (to mimic WIMP interactions) and 
photon sources
indicate that the larger ionization/recoil energy ratio of electron recoils results in
discrimination against photon backgrounds 
%better than 
$>10,000$:1 down to the 20\,keV threshold energy. 
In an exposure of $\sim$1\,kg of detectors for a few months, EDELWEISS detected 40~candidate events~\cite{edelweiss05}.  The six at high energies were consistent with the expected neutron background.  
As shown in Fig.~\ref{cryo}, most of the candidates were at low energies, likely caused by surface electron recoils, which result in incomplete charge collection and hence ionization/recoil ratios as low as those of nuclear recoils. 

The second phase of the EDELWEISS experiment is poised to begin.  A new, 120-detector cryostat has been successfully commissioned. After reducing electronics noise, the collaboration plans to install as much as 9\,kg of improved and cleaner detectors.  In addition to 21~detectors similar to those run previously, they will also run some metal-insulator transition detectors, which allow
discrimination of the dominant surface-event background by means of an additional fast signal for surface events.  They have also improved their shielding, both to reduce their neutron background and to attempt to reduce contamination near their detectors. 

The CRESST experiment also uses cryogenic detectors, but measures 
phonons and scintillation light, 
which provides as good rejection of surface events as it does rejection of photons in the bulk
due to the much larger light yield from all electron recoils relative to nuclear recoils.  
 A disadvantage of the small scintillation signal of nuclear recoils is that only high-energy WIMP events would produce any detectable light at all, resulting either in a high energy threshold, or in potential susceptibility that something causing phonons but no light may mimic a WIMP signal. 
CRESST ran two prototype detectors for a couple months in 2003~\cite{cresst}.  The run was without neutron shielding, so a significant neutron background on the oxygen in their CaWO$_4$ 
%(calcium tungstate) 
detectors was 
%expected and 
observed.
As shown in Fig.~\ref{cryo}, there were no candidate WIMP events on tungsten between 12--40\,keV.

The CRESST collaboration is nearing completion of major upgrades undertaken since taking this data.  They have installed neutron shielding and a muon veto, as well as new readout, electronics, and data acquisition sufficient for up to 33 detectors (10\,kg target mass).  
Eight detectors totaling 2.4\,kg were installed in September 2006 and should be run extensively through 2007.  
For the longer timescale, together with the EDELWEISS collaboration and others, they have formed a new collaboration, EURECA, dedicated to a cryogenic experiment at or near a ton of detector mass.

The CDMS~II experiment uses cryogenic Ge or Si detectors  that discriminate 
well 
against the otherwise dominant surface electron recoils by collecting 
fast, 
athermal phonons with thousands of thin-film sensors in addition to ionization.  
The ionization yield allows near-perfect separation of nuclear recoils from bulk electron recoils, and 
the shape, timing, and energy partition of the phonon pulses allow rejection of events occurring near the detector
surfaces.
This rejection works because the athermal phonons from electron recoils are faster than those from nuclear recoils, particularly if the electron recoils occur near a detector surface.  
Accepting only events with 
both 
slow phonon pulses and low ionization yield rejects nearly all of the surface events (over 99.5\% above 10\,keV)  while keeping over half of the nuclear-recoil events. 

\begin{figure}
  \includegraphics[height=.3\textheight]{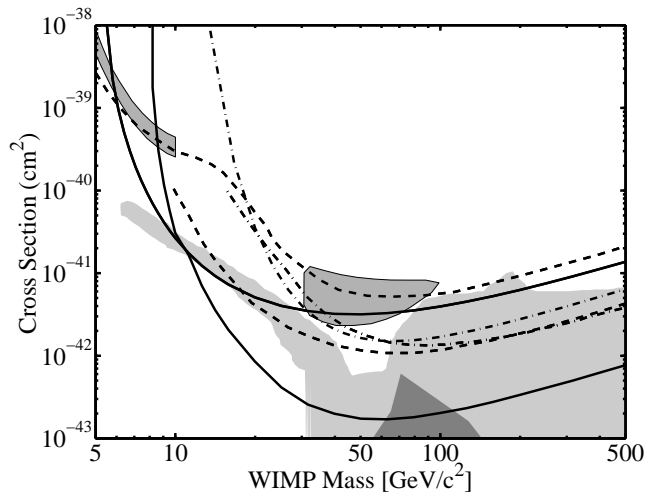}
  \caption{Current published upper limits  on the spin-independent WIMP-nucleon cross section 
  (90\% C.L.) versus WIMP mass under standard assumptions 
  about the WIMP distribution in the galaxy~\cite{lewinsmith}. 
  Upper limits based on Ge (lowest solid curve) and Si (upper solid curve except at low masses) detectors from CDMS~II at Soudan~\cite{r119prl} provide the best sensitivity for a wide range of WIMP masses, typically by nearly an order of magnitude, and exclude parts of the parameter space for minimal supersymmetric standard models (MSSM) without gaugino mass unification 
  (light grey shaded region)~\cite{Bottino03}, 
and for the constrained MSSM (dark shaded region)~\cite{ellis05}. 
The filled regions in the upper left~\cite{gondolo05} and center~\cite{DAMA} are consistent with the signal claimed by DAMA (1-4). 
Additional experimental upper limits are shown from DAMA~\cite{bernabei96}  (upper dashed curve), CRESST~\cite{cresst} (upper dashed-dot curve), EDELWEISS~\cite{edelweiss05} (lower dashed-dot curve), and ZEPLIN-I~\cite{zeplin05} (lower dashed curve). 
} \label{CurrentLimits}
\end{figure}

In 2004, 1.25\,kg of good Ge detectors and 0.4\,kg of good Si detectors were run for 74~days.
An analysis in which data-selection cuts were set ``blind,'' based on events from calibration sources or other events that could not be from WIMPs, 
resulted in no WIMP candidates in Si and one 
%WIMP candidate 
in Ge, consistent with the expected background~\cite{r119prl}.  
Figure~\ref{CurrentLimits} shows the resulting upper limits 
on the spin-independent WIMP-nucleon cross section under standard assumptions about how WIMPs are distributed in the galactic halo.  
CDMS limits are a factor of 6
lower than those of ZEPLIN-1
and an order of magnitude lower than those of EDELWEISS and CRESST. 
The results 
are clearly incompatible with the signal claimed by DAMA~\cite{DAMA} under the standard assumptions, 
and are also inconsistent with an alternate interpretation~\cite{gondolo05}
that the DAMA signal could be from low-mass WIMPs interacting not with DAMA's iodine but with their sodium, but some alternatives exist that could make the results compatible. 

CDMS is about to start taking more data, with 18 additional detectors for a total of almost 5\,kg of Ge. Running the detectors through 2007 should improve the experiment's sensitivity by another order of magnitude.  Further increases in mass are planned, first to a 25-kg phase, and eventually to a ton. The 25-kg phase would achieve zero background by increasing detector thickness from 1\,cm to 1\,inch, making additional modest improvements to the detectors and analysis, 
and siting the experiment deeper underground.

\subsection{Noble Liquid Detectors}

Detectors using noble liquids also show great promise for WIMP detection. 
A recoil in liquid Xe, for example,  induces both ionization and excitation of Xe atoms. The excitation produces photon emission by a fast singlet or by a slow triplet. Nuclear recoils result in fewer triplet decays and faster recombination, so nuclear recoils have a faster pulse shape than electron recoils. The effect is particularly pronounced in Ar and Ne, leading to extremely good discrimination in Ar and Neon based on timing alone.  
Additional discrimination is possible based on the relative amount of primary scintillation versus ionization.  This technology has the advantage of easier scaling to large masses since it is based on liquids.  The primary challenges are maintaining discrimination to low energies, since there are far fewer quanta than with the solid-state detectors, and overcoming radioactive backgrounds 
such as $^{85}$Kr in Xe or especially $^{39}$Ar in liquid Ar. 

ZEPLIN-1 is the lone completed liquid noble experiment. It was a single-phase experiment,
so it did not collect an ionization signal, but depended solely on pulse-shape discrimination, 
which in Xe is far from spectacular.  In 293\,kg-days exposure of 3.2\,kg fiducial mass, no excess consistent with nuclear recoils was seen~\cite{zeplin05}; however, the published limits are somewhat controversial because no {\it in situ} neutron calibration was performed~\cite{zeplinbad}.  
Three other single-phase experiments are in progress.  DEAP~\cite{deap2006}  and 
CLEAN~\cite{clean_ijmpa2005} both take advantage of the much larger timing difference between nuclear recoils and electron recoils in the lighter nobles argon and neon, while XMASS~\cite{xmass}, which uses Xe,
intends to take advantage of self-shielding to create a low-background 100-kg fiducial volume within
800\,kg of instrumented Xe.

WARP is the dual-phase experiment with the best current limits.  
Ionization electrons are drifted out of the liquid into gaseous argon, where they are detected by electroluminescence.  
Discrimination is 
based both on the shape of the primary scintillation
and on the ratio of primary scintillation to ionization.
The first running of a 3.2\,kg fiducial prototype 
resulted in no candidate events above 42\,keV in a 96.5 kg day exposure~\cite{warp}.  
Impressively, this prototype already produces 
competitive (preliminary) upper limits.
The collaboration is building a 140-kg experiment.  Due to the large background from $^{39}$Ar,
$10^7$:1 rejection will be needed;
already $4\times10^{5}$:1 rejection has been demonstrated.

Dual-phase Xe experiments, similar in design to WARP, are also in progress.  
ZEPLIN-II is operating 32\,kg of Xe and expects  
5000\,kg-days of exposure by summer 2007.  
%The zeplin program is slated to continue with Zeplin III, which seeks to improve over zeplin II by having their phototubes immersed in the xenon and by using a higher electric field, and one-ton versions of the technology called zeplin four or zeplin max.  
The XENON collaboration~\cite{xenon} 
started its first low-background run of a 15\,kg detector in August 2006. 
Plans exist for rapid increases in mass to ton-scale detectors.

\section{Conclusions and Prospects}

Currently, CDMS~II is the most sensitive WIMP-search experiment, with no sign of WIMPs
at spin-independent WIMP-nucleon cross sections near $10^{-7}$\,pb,  $\sim$6$\times$ lower than any other published experiment. 
%Incompatible with DAMA for scalar coherent interactions, standard halo
%tarting to probe mSUGRA region
However, the next year or so should see significant improvements from many experiments.
Comparable sensitivity is likely to be achieved soon by EDELWEISS, CRESST, ZEPLIN-II, XENON, and WARP, and all of the above experiments (including CDMS~II) may achieve $10\times$ improvements
by the end of 2007.

The next five years should see sensitivity improve to $10^{-9}$\,pb, or possibly even $10^{-10}$\,pb
if promising technologies can achieve the necessary background rejection. 
In particular, it is unclear what is the best technology for the ton scale needed to achieve  $10^{-10}$\,pb sensitivity.  It may be any one of the technologies described above, or it may be a new method.
As one example, %an inexpensive, 
a promising technology uses a superheated liquid or droplets as a threshold detector.  
By tuning thermodynamic parameters, the detector may be made insensitive to the low energy density deposited by a minimum-ionizing electron recoil.  Only a dense energy deposition, such as from a nuclear recoil, will provide enough energy to 
cause nucleation.  The COUPP collaboration appears to have met the principle challenge of keeping these detectors stable~\cite{coupp}.
The %primary 
attraction of these detectors is that they could allow inexpensive scaling to very large masses with  a broad range of materials and without need of cryogens or photon shielding.

A direct detection of WIMPs would warrant follow-up in order to learn as much as possible about them.
Measurement of the WIMP recoil spectrum with good statistics would constrain the WIMP's mass,
potentially demonstrating that a particle produced at accelerators indeed comprises the dark matter in the galaxy.  Simulations indicate that  as few as 10 events can place some interesting constraints on the WIMP's mass (although it is difficult to obtain an upper limit on the WIMP's mass, particularly for high-mass WIMPs), and a spectrum of 1000 events could result in $\sim$20\% uncertainty on the WIMP
mass, likely dominated by uncertainties on the WIMP distribution in the galactic halo.
Such large statistics (provided experimental operation is sufficiently stable)
could also take advantage of the expected annual modulation of the WIMP signal to 
confirm the extraterrestrial origin of the WIMPs and to learn more about their distribution in the galaxy.
In addition, detection using different target nuclei would potentially allow determination of both the spin-independent and spin-dependent cross sections.

A measure of the direction of the recoiling nucleus would provide additional information on
the distribution of WIMPs in the galaxy, allowing WIMP astronomy.
%. In the rest frame of the detector, most WIMPs (and the fastest ones) will be coming from the direction opposite the sun's direction of motion, so an experiment that determines the direction of recoils can actually perform WIMP astronomy.  
The most practical known way to determine the recoil direction is by drifting negative ions in a time projection chamber, so the tiny recoil distance can be recorded accurately.  
The DRIFT~\cite{drift} experiment has provided a proof of principle, but it remains to be seen
if the WIMP-nucleon cross section is large enough that one can build 
gas detectors with enough mass to detect the signal.  

Ultimately, the combination of WIMP direct and indirect detection with studies at colliders and
of the cosmic microwave background could answer fundamental questions beyond whether WIMPs are the dark matter and neutralinos exist.  
These combinations can determine whether the WIMPs are stable and whether there is non-baryonic dark matter other than WIMPs.
As thermal relics, the WIMPs could provide a window to the early universe, or we may learn that the WIMPs must have been generated out of thermal equilibrium.
WIMP astronomy could teach us about galaxy formation.
Furthermore, the combination of information from  these several methods may constrain 
the particle properties significantly better than colliders alone~\cite{baltz_lcc}.
Direct detection therefore should maintain excellent complementarity with the LHC and beyond.

\def\journal#1, #2, #3, #4#5#6#7{ % Journal reference.  Comma sets
  #1~{\bf #2}, #3 (#4#5#6#7)} % off: name, vol, page, year
\def\apl{\journal Appl.\ Phys.\ Lett., }
\def\apj{\journal Astrophys.\ J., }
\def\apjs{\journal Astrophys.\ J.\ Suppl., }
\def\app{\journal Astropart.\ Phys., }
\def\baas{\journal Bull.\ Am.\ Astron.\ Soc., }
\def\ejpc{\journal Eur.\ J.\ Phys.\ C., }
\def\lnp{\journal Lect.\ Notes\ Phys., }
\def\nature{\journal Nature, }
\def\nc{\journal Nuovo Cimento, }
\def\nima{\journal Nucl.\ Instr.\ Meth.\ A, }
\def\np{\journal Nucl.\ Phys., }
\def\npps{\journal Nucl.\ Phys.\ (Proc.\ Suppl.), }
\def\pl{\journal Phys.\ Lett., }
\def\prep{\journal Phys.\ Rep., }
\def\pr{\journal Phys.\ Rev., }
\def\prc{\journal Phys.\ Rev.\ C, }
\def\prd{\journal Phys.\ Rev.\ D, }
\def\prl{\journal Phys.\ Rev.\ Lett., }
\def\rsi{\journal Rev. Sci. Instr., }
\def\rpp{\journal Rep.\ Prog.\ Phys., }
\def\sjnp{\journal Sov.\ J.\ Nucl.\ Phys., }
\def\solarphys{\journal Solar Phys., }
\def\jetp{\journal J.\ Exp.\ Theor.\ Phys., }
\def\arnps{\journal Annu.\ Rev.\ Nucl.\ Part.\ Sci., }
%

%\begin{thebibliography}{99}

\end{document}